\documentclass[twoside]{article}

\renewcommand{\thefootnote}{\fnsymbol{footnote}}  
\usepackage{amssymb}
\usepackage{graphicx}
\usepackage{amsmath}
\usepackage{amsbsy}
\usepackage{epsfig}
\usepackage{amsthm}
\usepackage{bbm}
\usepackage{bm}
\newcommand{\id}{\mathbbm{1}}
\newcommand{\tr}{\textnormal{Tr}}
\newcommand{\beq}{\begin{eqnarray}}
\newcommand{\eeq}{\end{eqnarray}}
\newcommand{\bra}[1]{\ensuremath{\langle #1 |}}
\newcommand{\ket}[1]{\ensuremath{| #1 \rangle}}

\long\def\symbolfootnote[#1]#2{\begingroup%
\def\thefootnote{\fnsymbol{footnote}}\footnote[#1]{#2}\endgroup} 

\begin{document}
\setlength{\textheight}{8.0truein}    


\thispagestyle{empty}
\setcounter{page}{1}


\vspace*{0.88truein}



\vspace*{0.035truein}
\centerline{\bf CRITERION FOR $K$-SEPARABILITY IN MIXED MULTIPARTITE SYSTEMS}
\vspace*{0.37truein}
\centerline{\footnotesize
ANDREAS GABRIEL\symbolfootnote[1]{andreas.gabriel@univie.ac.at}, BEATRIX C. HIESMAYR\symbolfootnote[2]{beatrix.hiesmayr@univie.ac.at}, MARCUS HUBER\symbolfootnote[3]{marcus.huber@univie.ac.at}}
\vspace*{0.015truein}
\centerline{\footnotesize\it Faculty of Physics, University
of Vienna, Boltzmanngasse 5}
\baselineskip=10pt
\centerline{\footnotesize\it 1090 Vienna, Austria}
\vspace*{0.225truein} 
\vspace*{0.21truein}

Using a recently introduced framework, we derive criteria for quantum $k$-separability, which are very easily computed. In the case $k=2$, our criteria are equally strong to the best methods known so far, while in all other cases there are currently no comparable criteria known. We also show how the criteria can be implemented experimentally.
\vspace*{10pt}

\vspace*{3pt}


\section{Introduction}
Ever since quantum entanglement has started to be taken seriously by the scientific community, it has been subject to intensive studies and research. As soon as the properties of quantum mechanical systems were recognised not only to cause difficulties but also to allow for completely new applications and technology (which are far from comparison in classical physics, see e.g. Refs.~\cite{horodeckiqe, nagaj, brno, crypto}), entanglement played an important role in the rise of whole fields of research, such as quantum information theory.\\
\\
Although the nature of bipartite quantum entanglement is being more and more understood~\cite{horodeckiqe}, multipartite entanglement still holds many puzzling open questions. For example, although the concept of genuine multipartite entanglement is widely understood, tools for its detection are only beginning to be developed (see e.g. Refs.~\cite{guehnewit, guehnecrit, hmgh, brukner}). Since genuine multipartite entanglement and partial separability are very important concepts for various quantum informational tasks (e.g. quantum secret sharing~\cite{horodeckiqe}) and especially the former is believed to also play an important role in nature (as for example in photosynthesis~\cite{bio}), constructing such tools is highly necessary. In particular, detection of genuine $k$-nonseparability in mixed states is essential, since it is necessary in order to distill genuinely $k$-nonseparable pure states~\cite{distill}.\\
\\
This letter is organised as follows. In section~\ref{sec_defs}, the basic terminology and definitions are reviewed, such that in section~\ref{sec_crit} we can present our criterion for $k$-separability (which is the main result of this letter). This criterion, along with several examples of its application, are then discussed in section~\ref{sec_exs} as is its potential experimental realisation in section~\ref{sec_exp}.

\section{Motivation and Definitions}\label{sec_defs}
In Ref.~\cite{hmgh}, a general framework was introduced, which allows for construction of very versatile separability criteria. In this letter, we will use this framework do derive necessary criteria for $k$-separability of quantum states. Firstly, the basic terminology needs to be defined.\\

\begin{figure}[ht!]\centering\includegraphics[width=0.50\textwidth]{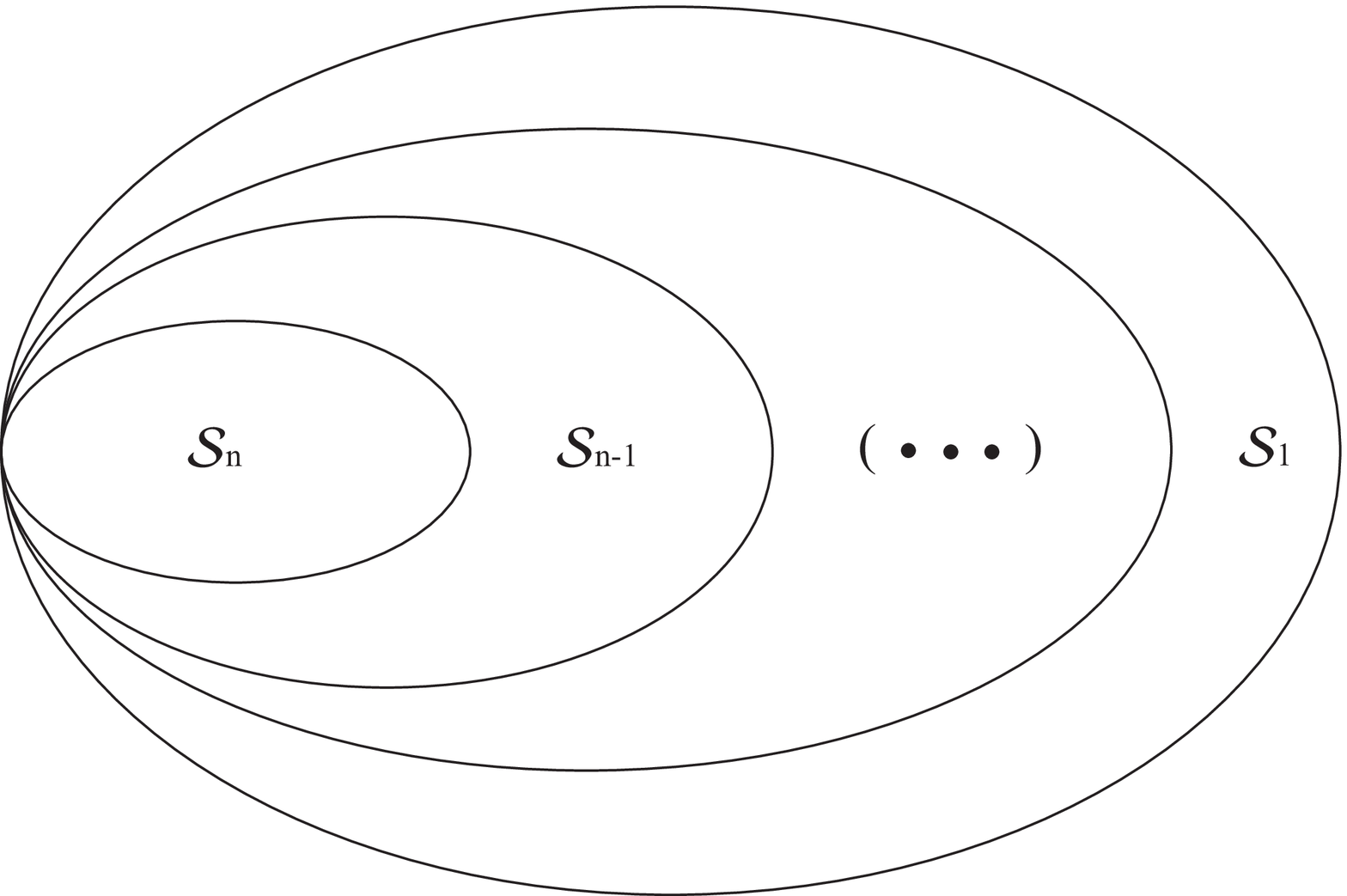}	\caption{Illustration of the geometry of the sets $S_k$, i.e. the sets of all $k$-separable states. Each set is convex and embedded within the next set: $S_n\subset S_{n-1}\subset\cdots\subset S_2\subset S_1$.}\label{fig_ksep}\end{figure}

\noindent\emph{Definition:} An $n$-partite pure quantum state $\left|\Psi_{k-sep}\right\rangle$ is called $k$-separable, iff it can be written as a product of $k$ substates:
\beq \left|\Psi_{k-sep}\right\rangle = \ket{\Psi_1}\otimes\ket{\Psi_2}\otimes \cdots \otimes \ket{\Psi_k} \eeq
A mixed state $\rho_{k-sep}$ is called $k$-separable, iff it has a decomposition into $k$-separable pure states:
\beq \rho_{k-sep} = \sum_{i} p_i \ket{\Psi_{k-sep}^i}\bra{\Psi_{k-sep}^i} \eeq
In particular, an $n$-partite state is called fully separable, iff it is $n$-separable. It is  called genuinely $n$-partite entangled, iff it is not biseparable (2-separable). Note that the individual pure states composing a $k$-separable mixed state may be $k$-separable under different partitions. Hence, in general, $k$-separable mixed states are not separable w.r.t. any specific partition, which makes $k$-separability rather difficult to detect. Let us also remark that whenever a state is $k$-separable, it is automatically also $k'$-separable for all $k' < k$ (as illustrated in Fig. \ref{fig_ksep}).\\
\\
In order to formulate our criterion for $k$-separability, we first need to define permutation operators $P_i$ acting on two copies of an $n$-partite state. These operators swap the $i$-th subsystems of the two copies:
\beq P_i \ket{\Psi_{a_1,a_2,\cdots,a_n}}\otimes\ket{\Psi_{b_1,b_2,\cdots,b_n}} = \ket{\Psi_{a_1,a_2,\cdots,a_{i-1},b_i,a_{i+1},\cdots,a_n}}\otimes\ket{\Psi_{b_1,b_2,\cdots,b_{i-1},a_i,b_{i+1},\cdots,b_n}} \eeq
where the $a_j$ and $b_j$ indicate the subsystems of the first and second copy of the state, respectively.\\
It is evident that if the permuted subsystem is separable from the rest of the state, the state is invariant under such a permutation. This observation is the key to constructing very general separability criteria.\\
In the following, permutation operators on sets of subsystems will be used, which is equivalent to applying several single permutation operators at once.

\section{Criterion for $k$-separability}\label{sec_crit}
\noindent Now, we can state our main result:\\
\emph{Theorem:} Every $k$-separable state $\rho$ satisfies
\begin{equation} \label{ksepineq}\tag{$\ast$}
\sqrt{\bra{\Phi}\rho^{\otimes 2}P_{tot}\ket{\Phi}}-\sum_{\{\alpha\}}\left(\prod_{i=1}^{k}\bra{\Phi}P^{\dagger}_{\alpha_i}\rho^{\otimes 2}P_{\alpha_i}\ket{\Phi}\right)^{\frac{1}{2k}}\leq 0
\end{equation}
for all fully separable states $\left|\Phi\right\rangle$, where the sum runs over all possible partitions $\alpha$ of the considered system into $k$ subsystems, the permutation operators $P_{\alpha_i}$ are the operators permuting the two copies of all subsystems contained in the $i$-th subset of the partition $\alpha$ and $P_{tot}$ is the total permutation operator, permuting the two copies.\\
\\
\emph{Proof:} To prove this, observe that (like in Refs.~\cite{guehnecrit,hmgh}) the the inequality is a convex function of $\rho$ (since the first term is the absolute value of a density matrix element and each term in the sum is the $2k$-th root of the product of $2k$ density matrix diagonal elements). Consequently, it suffices to prove the validity for pure states and validity for mixed states is guaranteed. So, let us assume w.l.o.g., that the given pure state $\rho$ is $k$-separable w.r.t. the $k$-partition $\tilde{\alpha}$. Due to its separability, $\rho$ is invariant under permutation of each element of $\tilde{\alpha}$:
\beq P_{\tilde{\alpha}_i}^\dagger \rho^{\otimes 2} P_{\tilde{\alpha}_{i}} = \rho^{\otimes 2} . \eeq
Therefore, the corresponding term in the sum can be written as
\beq \nonumber
\left(\prod_{i=1}^{k} \bra{\Phi} P_{\tilde{\alpha}_{i}}^\dagger \rho^{\otimes 2} P_{\tilde{\alpha}_{i}} \ket{\Phi}\right)^{\frac{1}{2k}} && = 
\left(\prod_{i=1}^{k} \bra{\Phi} \rho^{\otimes 2} \ket{\Phi}\right)^{\frac{1}{2k}} = \left(\prod_{i=1}^{k} \sqrt{\bra{\phi_1}\rho\ket{\phi_1}\bra{\phi_2}\rho\ket{\phi_2}} \right)^{\frac{1}{k}} = \\ && = \sqrt{\bra{\phi_1}\rho\ket{\phi_1}\bra{\phi_2}\rho\ket{\phi_2}} 
\eeq
where we used $\ket{\Phi}=\ket{\phi_1}\otimes\ket{\phi_2}$. Using $P_{tot} \ket{\phi_1}\otimes\ket{\phi_2} = \ket{\phi_2}\otimes\ket{\phi_1}$, we can now rewrite ineq. (\ref{ksepineq}) as
\beq \left|\bra{\phi_1} \rho \ket{\phi_2}\right| - \sqrt{\bra{\phi_1}\rho\ket{\phi_1}\bra{\phi_2}\rho\ket{\phi_2}} - \sum_{\{\alpha\neq\tilde{\alpha}\}}\left(\prod_{i=1}^{k} \bra{\Phi} P_{\alpha_i}^\dagger \rho^{\otimes 2} P_{\alpha_i} \ket{\Phi}\right)^{\frac{1}{2k}} \leq 0 \eeq
Now, the first term is an off-diagonal matrix-element and the second term is the squareroot of the product of the two corresponding diagonal elements, hence (and because $\rho$ is a pure state), the first two terms cancel each other. It is evident that the remaining sum over strictly nonnegative terms (diagonal elements) with a negative sign is nonpositive, which proves our theorem.\qed

\section{Results and Discussion}\label{sec_exs}
\subsection{Example 1 - Three QuBits and Three QuTrits}
\begin{figure}[ht!]\centering\includegraphics[width=7cm,keepaspectratio=true]{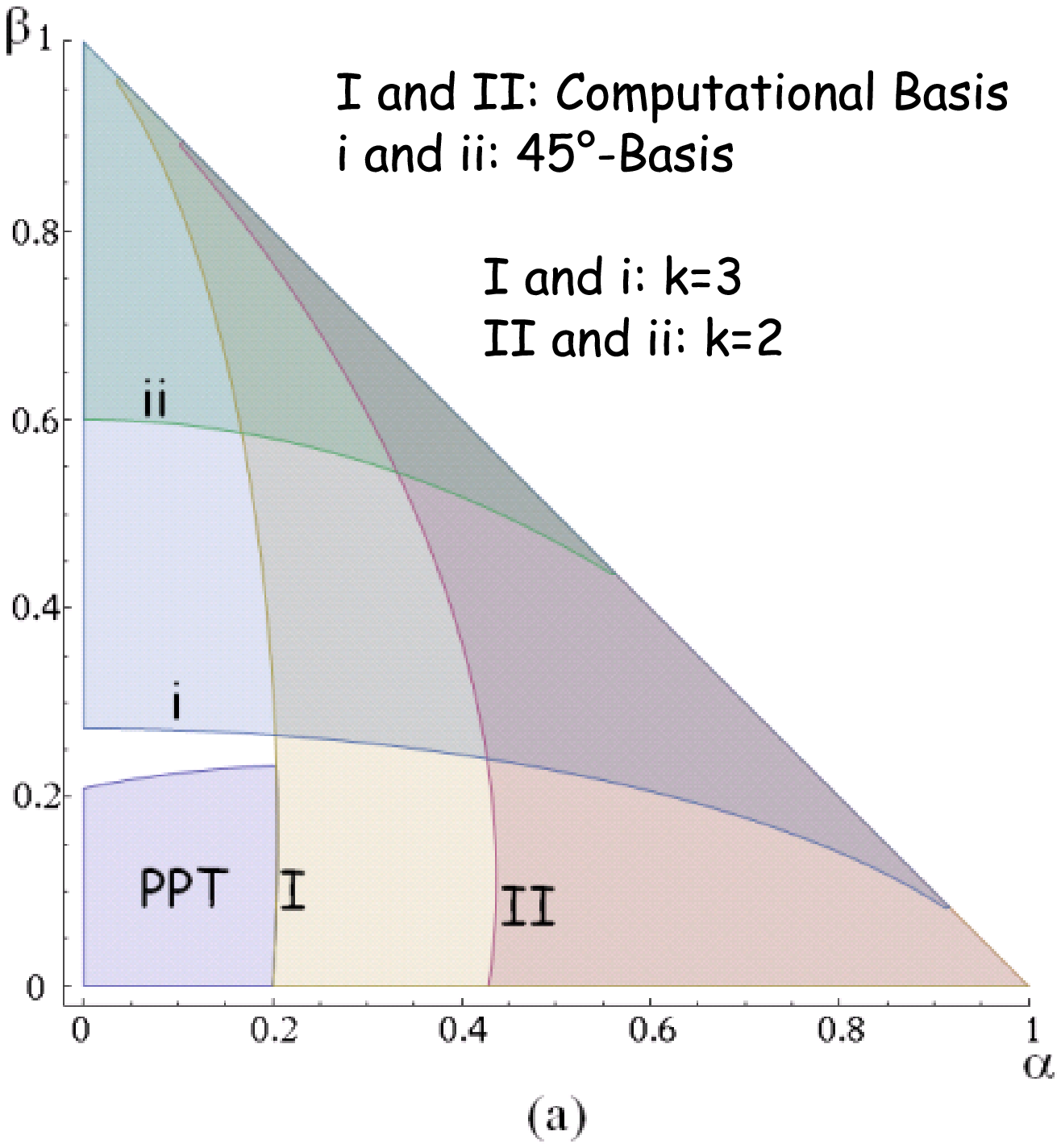}\includegraphics[width=7cm,keepaspectratio=true]{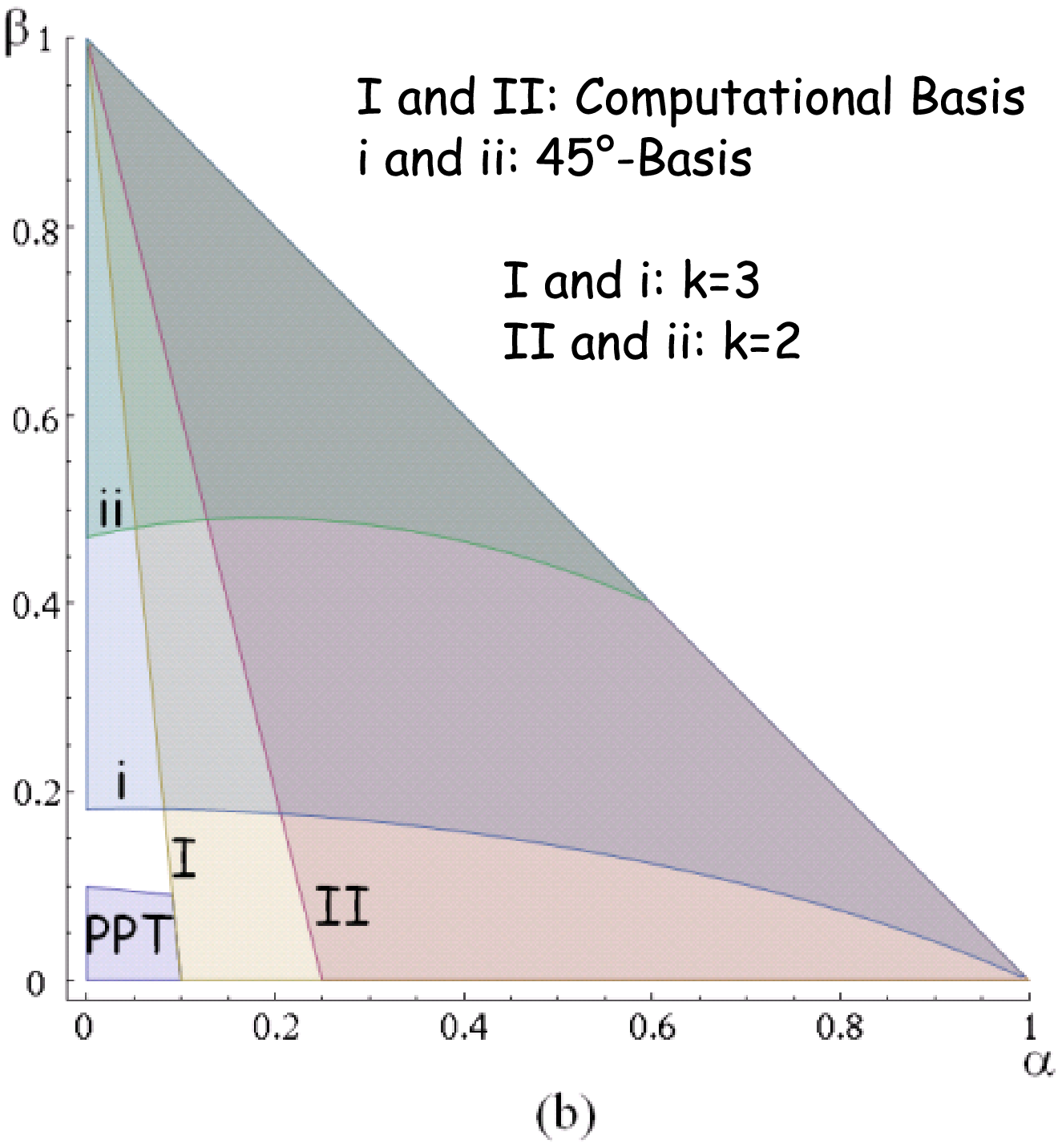}\caption{(Colour online): Illustration of the detection parameters of (a) the three-qubit-state~(\ref{3qubitstate}) and (b) the three-qutrit-state~(\ref{3qutritstate}) for the inequality~(\ref{ksepineq}). The lines I and II represent the thresholds of detection for entanglement~($k=3$) and genuine tripartite entanglement~($k=2$) in the computational basis. The lines i and ii represent the same in the basis rotated by (a) $\pi/4$ and (b) $\pi/4$ around two axes.}\label{fig_ex1}\end{figure}

Consider the family of three-qubit-states
\beq \label{3qubitstate} \rho = \alpha \ket{GHZ}\bra{GHZ} + \beta \ket{W}\bra{W} + \frac{1-\alpha-\beta}{8} \id \eeq
with the well-known $GHZ$-state $\ket{GHZ} = (\ket{000} + \ket{111})/\sqrt{2}$ and $W$-state $\ket{W}=(\ket{001}+\ket{010}+\ket{100})/\sqrt{3}$.\\
It turns out that for states in the vicinity of the $GHZ$-state, $\ket{\Phi} = \ket{000111}$ is a good choice, while the $W$-state is not detected at all in the computational basis. On the other hand, the $W$ state can be detected in a basis rotated by $\pi/4$, where in turn the $GHZ$-state is not detected. In Fig.~\ref{fig_ex1}(a) the detection quality of ineq. (\ref{ksepineq}) is illustrated for both these choices for $\ket{\Phi}$.\\

\noindent Consider now the following generalisation of these states to three-qutrit-systems:
\beq \label{3qutritstate} \rho = \alpha \ket{gGHZ}\bra{gGHZ} + \beta \ket{\xi}\bra{\xi} + \frac{1-\alpha-\beta}{27} \id \eeq
with the generalised $GHZ$-state $\ket{gGHZ} = (\ket{000}+\ket{111}+\ket{222})/\sqrt{3}$ and the state $\ket{\xi} = (\ket{012}+\ket{021}+\ket{102}+\ket{120}+\ket{201}+\ket{210})/\sqrt{6}$, which is a possible generalisation of the $W$-state (as it also is genuinely 3-partite entangled while having entangled reduced density matrices). Concerning the choice of $\ket{\Phi}$, the same arguments as above apply and Fig.~\ref{fig_ex1}(b) shows the detection quality of our criterion in this case (where the rotated basis is rotated around the $y$- and $z$-axis by $\pi/4$ each).

\subsection{Example 2 - General $GHZ$-States}
The most general maximally entangled state for $n$ qudits has the form
\beq \ket{\Psi}=\sum_{i=0}^{d-1} \ket{i}^{\otimes n} . \eeq
Consider this state dampened by isotropic noise, i.e. the family
\beq \rho = p \ket{\Psi}\bra{\Psi} + \frac{1-p}{d^n} \id_{d^n} \eeq
Straightforward computation of the criterion (\ref{ksepineq}) yields that these states are not $k$-separable if
\beq p > \frac{\gamma}{\gamma + d^{n-1}} \eeq
where $\gamma$ is the number of possible partitions, i.e. depends only on $k$ and $n$.\\
Note that for general entanglement detection~($k=n$), this criterion is as strong as the PPT criterion, since $\gamma=1$ and thus $\rho$ is detected to be entangled for $p > (1+d^{n-1})^{-1}$, which is exactly the threshold detected by the PPT criterion.

\subsection{Example 3 - Spin Chain}
A multipartite quantum system cannot exhibit arbitrary entanglement
properties. This has several implications in the properties of
spin chains and spin lattices, which are the typical subjects of statistical
and solid state physics (see e.g. the review article~\cite{Amico}). As an example, let us consider translationally
invariant states of an infinite one dimensional chain of qubits
\cite{Poulsen, Meyer, HKN}, also known as finitely correlated states,
which are ground states of Hamiltonians, and intensively studied in
literature.\\
In Ref~\cite{HKN}, the authors investigated the maximally possible
achievable entanglement of nearest neighbours, which serves as a
reference point for interpreting entanglement values obtained for
real physical systems and sheds light onto the subtle properties of
spin-spin interactions. An $n$-qubit state of an infinite
translationally invariant spin-$\frac{1}{2}$ chain is given by
\begin{eqnarray}
\rho_{[1,2,\dots,n]}=\sum_{\mathbf{s},\mathbf{t}}
|\mathbf{s}\rangle\langle \mathbf{t}|
\tr(v_{\mathbf{s}}^\dagger\;\rho_B\; v_\mathbf{t})
\end{eqnarray}
with $|\mathbf{s}\rangle=|s_1\dots s_n\rangle$,
$v_{\mathbf{s}}=v_{s_1}\cdot v_{s_n}$ and $s_i=0,1$. Here the
operators $v_0,v_1$ and the auxiliary state $\rho_B$ are of
dimension $b$ and have to satisfy certain equations. They simulate
the ``rest of the chain'' and, consequently, the dimensionality $b$
approximates the qubit, two--qubit, \dots, $n$--qubit reduced states of the
pure state of the infinite qubit chain accordantly (i.e. the higher $b$, the better the approximation).\\
We want to find out whether the three--, four-- or five--qubit
states of this infinite chain with maximal nearest neighbour entanglement
are $k$-separable or for which $k$ our inequality is violated.
For that, we used the values given in Ref.~\cite{HKN} for the
case of maximised nearest neighbour entanglement for dimension
$b=2$ and $b=6$ (which was shown to be a good approximation). We
find that for all $k\leq n$ the inequalities (\ref{ksepineq}) are
violated, and thus the chain is not $k$-separable for any $k$, but genuinely multipartite entangled.

\subsection{Discussion}
It is quite clear from the above examples, that the introduced criteria are not only the first criteria for genuine $k$-nonseparability, but also work very well as criteria for detecting general quantum entanglement.\\
Although a $k$-separable state is always also $k'$-separable for all $k'<k$, our criteria do not include one another, i.e. if a state violates inequality~(\ref{ksepineq}) for a certain $k_1$, there may still be a certain $k_2>k_1$ for which the inequality is satisfied, even though the state is evidently not $k_2$-separable. It is unclear wether this can be exploited to improve the criterion (e.g. by using that the $m$-fold tensor product of a $k$-separable state is $(k \times m)$-separable).\\
\\
All our achieved results can be improved by optimising the used state $\ket{\Phi}$ over all local unitary transformations, i.e. transformations of the form $\rho \rightarrow U \rho U^{\dagger}$, where $U = U_1\otimes U_2\otimes \cdots \otimes U_n$. Although such an optimisation is a highly complex task, it can be done comperatively easily by means of the method described in Ref.~\cite{shh}. In many cases however, this is unnecessary, since our criteria are quite strong even without optimisation.

\section{Experimental implementation}\label{sec_exp}
Since a full quantum state tomography requires a huge number of measurements, it is important for separability criteria of multipartite systems to be experimentally implementable without such a procedure. With inequality~(\ref{ksepineq}) this is possible, due to the fact that it can be expressed in terms of density matrix elements, each of which can be measured efficiently with a single observable. For any fixed $\ket{\Phi}$, the number of density matrix elements in ineq.~(\ref{ksepineq}) is $2^n-1$, which -- compared to the $d^{2n}/2$ measurements needed for a full tomography -- not only grows significantly slower with $n$, but more importantly has the vast advantage of being independent of the dimensionality $d$ of the subsystems.\\
The observables follow directly from the choice of $|\Phi\rangle$. For the observables associated with the second term in ineq.~(\ref{ksepineq}), one only needs to measure diagonal density matrix elements
\begin{eqnarray}
\mathcal{O}_{\alpha_i}^{A/B}=|\chi_{\alpha_i}^{A/B}\rangle\langle\chi_{\alpha_i}^{A/B}| ,
\end{eqnarray}
where
\begin{eqnarray}
|\chi_{\alpha_i}^{A}\rangle\otimes|\chi_{\alpha_i}^{B}\rangle=P_{\alpha_i}|\Phi\rangle
\end{eqnarray}
and therefore
\begin{eqnarray}
\langle \mathcal{O}_{\alpha_i}^{A}\otimes\mathcal{O}_{\alpha_i}^{B}\rangle=\bra{\Phi}P^{\dagger}_{\alpha_i}\rho^{\otimes 2}P_{\alpha_i}\ket{\Phi} .
\end{eqnarray}
From the separability of $|\Phi\rangle$ follows the product form of the observables
\begin{eqnarray}
\mathcal{O}_{\alpha_i}^{A/B}=\bigotimes_{j=1}^n O_j .
\end{eqnarray}
These observables can be implemented by means of local observables $O_j$ of the subsystems. For the positive term of the inequality, one needs to measure a single off-diagonal density matrix element, which is also achievable via a single multipartite coherence measurement. An efficient way to do so is to measure the $n$-party visibility (see Refs.~\cite{Bergou,HH2,bramon}), which can be done using only local observables.

\section{Conclusion}
We have developed an easily computable criterion for detecting $k$-nonseparability in mixed states for arbitrary quantum systems and arbitrary values of $k$, which does not involve eigenvalue computation and works well even without numerical optimisation. In cases where comparable results exist (i.e. for the special case of detecting genuine multipartite entanglement, $k=2$) it is as good as the best known methods so far (i.e. the methods presented in \cite{guehnecrit} and \cite{hmgh}). Also, for the case $k=n$, i.e. general detection of entanglement, in many cases it is as strong as the PPT-criterion. We also discussed possibilities to implement our criterion experimentally.

\section{Acknowledgements}
\noindent The authors would like to thank Theodor Adaktylos, Florian Hipp, Christoph Spengler and Heidi Waldner for productive discussions and especially Hans Schimpf for his valuable assistance. Marcus Huber and Andreas Gabriel gratefully acknowledge the Austrian Fund project FWF-P21947N16.


\end{document}